# OPTICAL CHARACTERIZATION OF ORGANIC LIGHT-EMITTING THIN FILMS IN THE ULTRAVIOLET AND VISIBLE SPECTRAL RANGES

R.M. MONTEREALI, M.A. VINCENTI

ENEA - Sezione Sorgenti Laser e Acceleratori
Centro Ricerche Frascati, Roma

E. NICHELATTI

ENEA - Gruppo Dispositivi Ottici
Centro Ricerche Casaccia, Roma

F. DI POMPEO, E. SEGRETO, N. CANCI, F. CAVANNA

INFN - Laboratori Nazionali del Gran Sasso, L'Aquila e
Dipartimento di Fisica, Università dell'Aquila, Coppito (AQ)

AGENZIA NAZIONALE PER LE NUOVE TECNOLOGIE,
L'ENERGIA E LO SVILUPPO ECONOMICO SOSTENIBILE

# OPTICAL CHARACTERIZATION OF ORGANIC LIGHT-EMITTING THIN FILMS IN THE ULTRAVIOLET AND VISIBLE SPECTRAL RANGES

R.M. MONTEREALI, M.A. VINCENTI

ENEA - Sezione Sorgenti Laser e Acceleratori
Centro Ricerche Frascati, Roma

E. NICHELATTI

ENEA - Gruppo Dispositivi Ottici
Centro Ricerche Casaccia, Roma

F. DI POMPEO, E. SEGRETO, N. CANCI, F. CAVANNA

INFN - Laboratori Nazionali del Gran Sasso, L'Aquila e
Dipartimento di Fisica, Università dell'Aquila, Coppito (AQ)

## CARATTERIZZAZIONE OTTICA DI FILM SOTTILI ORGANICI EMETTITORI DI LUCE NELL'ULTRAVIOLETTO E NEL VISIBILE

R.M. MONTEREALI, M.A. VINCENTI, E. NICHELATTI, F. DI POMPEO, E. SEGRETO, N. CANCI, F. CAVANNA

**Riassunto**
La caratterizzazione spettrofotometrica di campioni otticamente attivi di elevata efficienza, ad esempio lastre e film sottili organici emettitori di luce, può risultare problematica poiché la loro luminescenza è catturata insieme ai segnali monocromatici trasmessi e riflessi e le conseguenti misure di trasmissione e riflessione ottica risultano perturbate alle lunghezze d'onda interne alla banda di fotoeccitazione. Infatti, nella maggior parte degli spettrofotometri commerciali, il fascio luminoso che incide sul campione è filtrato solo a monte di esso e non è ulteriormente filtrato prima di raggiungere il rivelatore. In questo Rapporto, presentiamo e discutiamo un metodo sviluppato per correggere spettri fotometrici perturbati da fotoluminescenza.

## *OPTICAL CHARACTERIZATION OF ORGANIC LIGHT-EMITTING THIN FILMS IN THE ULTRAVIOLET AND VISIBLE SPECTRAL RANGES*

***Abstract***
*The spectrophotometric characterization of high efficiency, optically-active samples such as light-emitting organic bulks and thin films can be problematic because their broad-band luminescence is detected together with the monochromatic transmitted and reflected signals, hence perturbing measurements of optical transmittance and reflectance at wavelengths within the photoexcitation band. As a matter of fact, most commercial spectrophotometers apply spectral filtering before the light beam reaches the sample, not after it. In this Report, we introduce and discuss the method we have developed to correct photometric spectra that are perturbed by photoluminescence.*

# INDEX

# OPTICAL CHARACTERIZATION OF ORGANIC LIGHT-EMITTING THIN FILMS IN THE ULTRAVIOLET AND VISIBLE SPECTRAL RANGES

## 1 INTRODUCTION

Commercial monochromator-based spectrophotometers are such that light is spectrally filtered before it reaches the sample, both in transmittance and reflectance mode, by a monochromator [1]. Between the sample and the instrument detector no further monochromator is present, so that the detector collects all the radiation coming from the sample within its acceptance angle. If the sample is optically active, it can happen that for certain incident wavelengths not only the sample transmits (reflects) light at the same incident wavelengths and according to its transmittance (reflectance) coefficients, but also that it emits broad-band luminescence at lower frequencies. In such a case, the spectrophotometer detector collects the transmitted (reflected) radiation together with photoluminescence (PL), thus recording a perturbed measurement. The perturbation can be relevant, especially if the measurements are performed with an integrating sphere. As a matter of fact, the released PL usually spreads over solid angles much wider than those of the transmitted (reflected) light, almost entirely captured by the sphere. On the other hand, in absence of an integrating sphere, the amount of PL within the narrower detection angle can be very small and hence neglected.

In principle, a low-pass filter interposed between sample and detector can help in diminishing the perturbation due to PL. In this Report, we formally analyze what the effect of such a filter is and how measured spectral data can be elaborated to try removing the perturbation and infer a more realistic spectrum.

## 2 THEORY

Let us assume that the optical transmittance and reflectance of a sample have to be measured in a spectrophotometer, and that optically-active centres are embedded in the sample. Let $\sigma_{PE}(\lambda)$ and $\sigma_{PL}(\lambda)$ be the photoexcitation (PE) and PL cross sections, respectively, of the centres, where $\lambda$ is the wavelength. Let $T_S(\lambda)$ and $R_S(\lambda)$ be the *true* optical intensity transmittance and reflectance, respectively, of the sample. For what discussed in the Introduction, the *measured* transmittance and reflectance are perturbed by PL in the following way:

$$T(\lambda) = T_S(\lambda) + \frac{\Omega_T}{4\pi} N_T \sigma_{PE}(\lambda) \int_{\Lambda_{PL}} \Gamma_T(\lambda')\sigma_{PL}(\lambda')\, d\lambda', \tag{1}$$

$$R(\lambda) = R_S(\lambda) + \frac{\Omega_R}{4\pi} N_R \sigma_{PE}(\lambda) \int_{\Lambda_{PL}} \Gamma_R(\lambda')\sigma_{PL}(\lambda')\, d\lambda'. \tag{2}$$

In these equations, $\Omega_T$ ($\Omega_R$) is the acceptance solid angle of the detector for the transmittance (reflectance) measurement, $N_T$ ($N_R$) is the number of active centres that interact with the sample beam in the transmittance (reflectance) measurement process, $\Gamma_T$ ($\Gamma_R$) is the spectral response of the sample-detector setup used to measure transmittance (reflectance) that depends on several parameters, such as sample position, detector spectral response, detection solid angle, etc. Finally, the spectral domain of integration, $\Lambda_{PL}$, is the wavelength range within which the PL cross section cannot be neglected. Obviously, the PL-induced perturbation is effective when the wavelength of the impinging sample beam lies within the excitation band. For other wavelengths, the perturbation terms in Eqs. (1) and (2) are null.

By inserting a suitable low-pass filter between sample and detector, one can try to minimize the unwanted contributions to the measured signal due to PL. The optical transmittance of such a filter, $T_F(\lambda)$, is required – for reasons that will become clear later – to have an almost constant, small value in the domain $\Lambda_{PL}$, and much larger values outside $\Lambda_{PL}$, that is

$$T_F(\lambda) \approx \varepsilon \text{ for } \lambda \in \Lambda_{PL} \quad \text{and} \quad T_F(\lambda) >> \varepsilon \text{ for } \lambda \notin \Lambda_{PL}. \tag{3}$$

Typically, a low-pass filter features a cut-off wavelength, $\lambda_0$. Ideally, for wavelengths longer than $\lambda_0$ the filter transmittance is almost zero, while for wavelengths shorter than $\lambda_0$ it is as high as the design allows. In the present case, it is important that $\lambda_0$ is shorter than the PL-band wavelengths, or at least that only a very small fraction of the PL band falls below $\lambda_0$, otherwise the correction we are going to illustrate could not satisfactorily work because the result would still include a residual fraction of perturbation due to PL. On the other hand, since no data correction can be applied at wavelengths longer than $\lambda_0$ because – as we will see later – the correction itself and/or its uncertainty would diverge there, one has to consider that, if $\lambda_0$ were *much* shorter than the bottom extreme of the PL-band, in a wide spectral interval the true optical properties of the sample would remain unknown and could be only guessed with data interpolation. An ideal filter should have a cut-off wavelength as close as possible to the shorter wavelength of the PL band. Figure 1 shows a scheme of the wavelength zones involved in the measurement and correction process.

By neglecting light multireflections occurring between sample and filter, the *measured* transmittance in presence of the filter is

$$\tilde{T}(\lambda) = T_F(\lambda)T_S(\lambda) + \frac{\Omega_T}{4\pi} N_T \sigma_{PE}(\lambda) \int_{\Lambda_{PL}} \Gamma_T(\lambda')T_F(\lambda')\sigma_{PL}(\lambda')\, d\lambda', \qquad (4)$$

which, thanks to the required conditions for the filter, can be approximated as

$$\tilde{T}(\lambda) \approx T_F(\lambda)T_S(\lambda) + \varepsilon \frac{\Omega_T}{4\pi} N_T \sigma_{PE}(\lambda) \int_{\Lambda_{PL}} \Gamma_T(\lambda')\sigma_{PL}(\lambda')\, d\lambda'. \qquad (5)$$

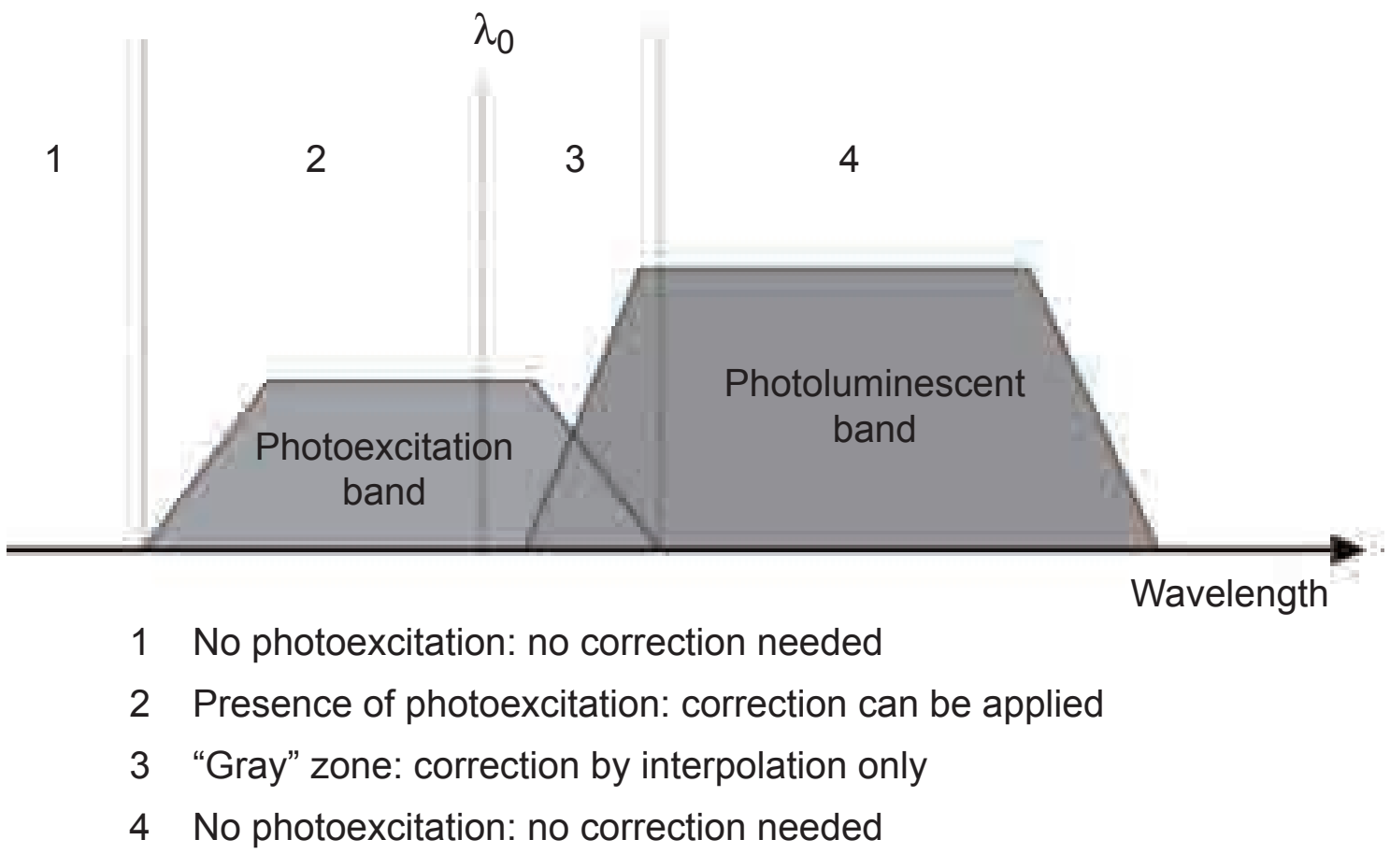

*Figure 1. Spectral zones for the optical characterization of a light-emitting sample and the correction of its PL-altered photometric spectra by means of a low-pass filter having cut-off wavelength $\lambda_0$. The correction procedure is described in detail in the text.*

If $\varepsilon$ is small enough that the second term in the right hand side of Eq. (5) can be neglected, the *true* sample transmittance is retrieved as

$$T_S(\lambda) \approx \frac{\tilde{T}(\lambda)}{T_F(\lambda)} \quad \text{for} \quad \lambda < \lambda_0. \tag{6}$$

However, in case the second term in the right hand side of Eq. (5) cannot be easily quantified, $T_S(\lambda)$ can be derived more accurately, without the above assumption on $\varepsilon$, from Eqs. (1) and (5). The result is

$$T_S(\lambda) \approx \frac{\tilde{T}(\lambda) - \varepsilon T(\lambda)}{T_F(\lambda) - \varepsilon} \quad \text{for} \quad \lambda < \lambda_0. \tag{7}$$

The wavelength range enclosed between $\lambda_0$ and the lower limit of the PL-band is a "gray" zone of the spectrum (see Fig. 1) where no correction can be applied, because therein the corrected value diverges being $T_F(\lambda) \approx \varepsilon$, thus transmittance can be only roughly estimated via interpolation. In the other parts of the spectrum, that is, outside the PE band, the direct measurements are assumed to be reliable because in principle no perturbation is present.

A similar derivation can be done for reflectance. The reflectance *measured* when the low-pass filter is interposed between sample and detector is

$$\tilde{R}(\lambda) \approx R_F(\lambda) + T_F^2(\lambda) R_S(\lambda) + \varepsilon \frac{\Omega_R}{4\pi} T_F(\lambda) N_R \sigma_{PE}(\lambda) \int_{\Lambda_{PL}} \Gamma_R(\lambda') \sigma_{PL}(\lambda')\, d\lambda', \tag{8}$$

where $R_F(\lambda)$ is the reflectance of the low-pass filter. Note the presence of the squared $T_F^2(\lambda)$ term due to the transmission of both the impinging and reflected beams through the filter. Here, too, if $\varepsilon$ is small enough that the third term in the right hand side of Eq. (8) can be neglected, on gets

$$R_S(\lambda) \approx \frac{\tilde{R}(\lambda) - R_F(\lambda)}{T_F^2(\lambda)} \quad \text{for} \quad \lambda < \lambda_0. \tag{9}$$

The corresponding more accurate solution is

$$R_S(\lambda) \approx \frac{\tilde{R}(\lambda) - R_F(\lambda) - \varepsilon T_F(\lambda) R(\lambda)}{T_F(\lambda)\left[T_F(\lambda) - \varepsilon\right]} \quad \text{for} \quad \lambda < \lambda_0 . \tag{10}$$

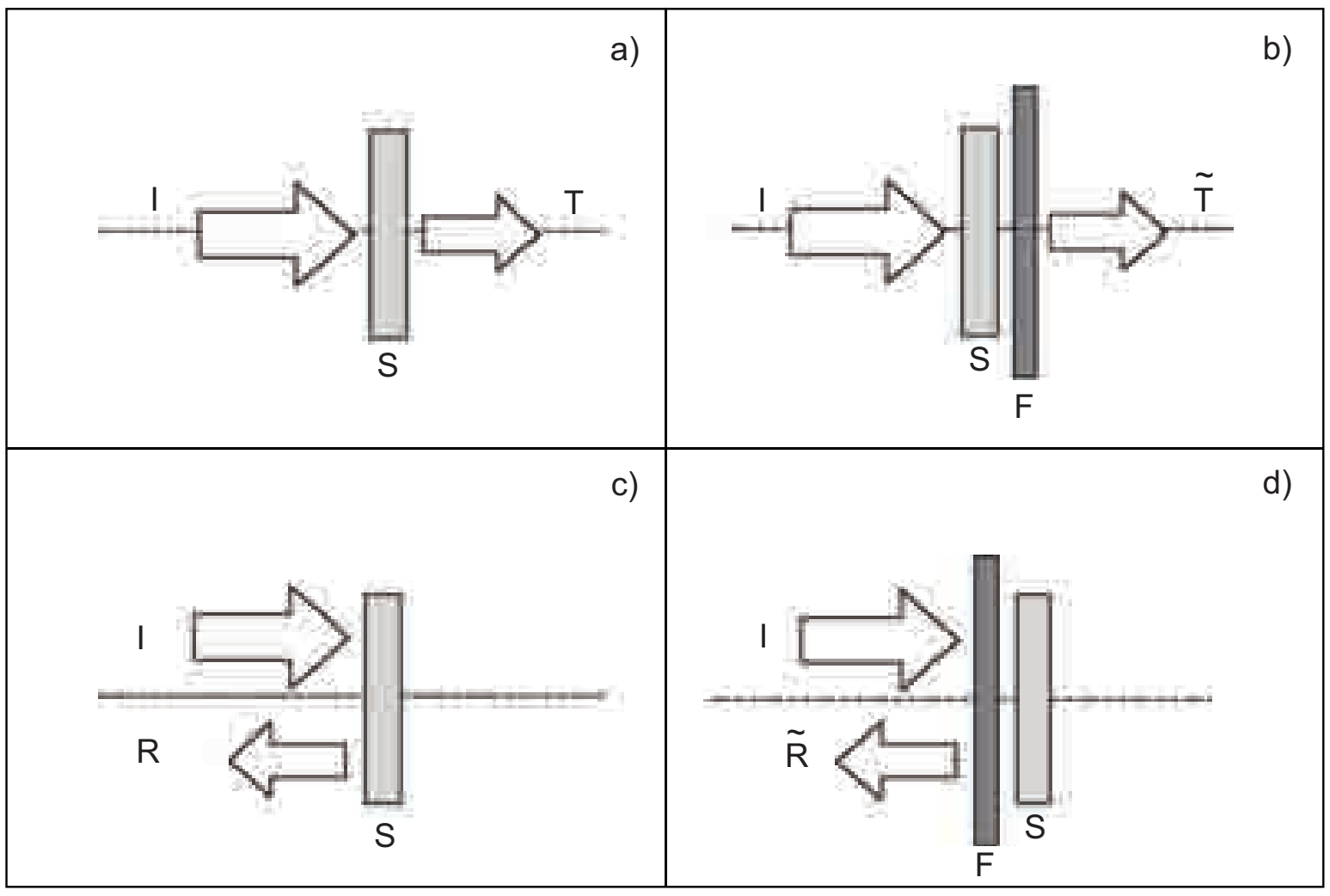

*Figure 2. Placement of sample, S, and filter, F, in a-b) transmittance and c-d) reflectance measurements. In the figure,* I *is the impinging collimated sample-beam;* T *and* R *are the transmitted and reflected beams, respectively, containing unwanted PL contributions – see Eqs. (1) and (5);* $\tilde{T}$ *and* $\tilde{R}$ *are the transmitted and reflected beams, respectively, in presence of the low-pass filter – see Eqs. (2) and (8).*

Again, note how this corrected value diverges for $\lambda \geq \lambda_0$.

Figure 2 schematically illustrates the mutual placement of sample and filter in the above-mentioned transmittance and reflectance measurements. The results found, mainly summarized by Eqs. (7) and (10), suggest an operative schedule to the spectrophotometric characterization and data correction of an optically-active sample.

1. The optical transmittance and reflectance of the sample, $T(\lambda)$ and $R(\lambda)$, are measured with a spectrophotometer. These measured spectra are possibly perturbed by PL at wavelengths within the spectral interval where the active centres of the sample are excited (PE band). Elsewhere, the measured spectra are assumed to be reliable.
2. The PE and PL spectra of the sample are measured by means of an instrument such as a spectrofluorimeter. This allows locating the PE and PL bands, thus estimating in which spectral zones the light-emitting properties of the sample could have perturbed the previously measured photometric spectra and in which zones a correction is possible.

3. From the results of the previous step, a suitable optical filter is selected and optically characterized to get its transmittance and reflectance spectral properties, $T_F(\lambda)$ and $R_F(\lambda)$ (including an estimation of $\varepsilon$ ).
4. The transmittance and reflectance of the sample are measured again, this time inserting the optical filter between sample and detector, obtaining $\tilde{T}(\lambda)$ and $\tilde{R}(\lambda)$. The best placement of the filter is in contact with the sample, so that the measurements are affected as poorly as possible by the increased thickness of the filter-sample assembly – this is important especially if the sample is of scattering nature and measurements are being taken with an integrating sphere.
5. The measured spectra are elaborated according to the previous theory to estimate the true spectra of the sample, $T_S(\lambda)$ and $R_S(\lambda)$, where applicable.

### 2.1 Error propagation

Understanding how effective this data-correction method can be is quite important. As a matter of fact, the illustrated theory is approximate – for instance, PL and PE band limits are surely not as sharp as the theory assumes, and the same is true for the transmittance spectrum of the low-pass filter. Moreover, the elaboration of the data introduces experimental error propagation that can become relevant in some parts of the spectrum. Additional errors are introduced if the selected filter does not exactly comply with the requests of Eqs. (3).

Indicating with the prefix $\delta$ the experimental error associated with each spectral quantity, one can verify that application of the logarithmic differentiation [2] to Eq. (7) brings

$$\frac{\delta T_S}{T_S} \approx \frac{\delta\tilde{T} + T\,\delta\varepsilon}{\left|\tilde{T} - \varepsilon T\right|} + \frac{\delta T_F + \delta\varepsilon}{T_F - \varepsilon}, \tag{11}$$

where we have omitted $\varepsilon\,\delta T$ in the numerator of the first right-hand side term because it is likely to be much smaller than both $\delta\tilde{T}$ and $T\,\delta\varepsilon$ . Note how the above uncertainty diverges at wavelengths longer than $\lambda_0$, where $T_F \approx \varepsilon$ . A similar expression holds for the uncertainty of the sample reflectance, see Eq. (19),

$$\frac{\delta R_S}{R_S} \approx \frac{\delta\tilde{R} + \delta R_F + T_F R\,\delta\varepsilon}{\left|\tilde{R} - R_F - \varepsilon T_F R\right|} + \frac{2\delta T_F + \delta\varepsilon}{T_F - \varepsilon}. \tag{12}$$

Here too, the smallest quantities have been neglected to simplify the result.

Equations (11) and (12) can be used to estimate the uncertainties associated with the corrected spectra. Regarding $\varepsilon$ and its associated uncertainty, if the transmittance of the filter is not spectrally uniform for $\lambda > \lambda_0$, one can assume $\varepsilon$ to be equal to its average value and $\delta\varepsilon$ to be given by the standard deviation of it. Also, if $\varepsilon$ is so small to be instrumentally zero, one will set $\varepsilon = 0$ and $\delta\varepsilon$ equal to the instrumental sensitivity. As far as Eq. (12) is concerned, one can notice that if the sample *true* reflectance is very small, the reflectance of the sample-filter assembly is very close to the reflectance of the filter alone, thus making very small the denominator of the first fraction in the right-hand side member and bringing a considerable uncertainty on the correction of $R_S$. All these considerations have to be taken into account when applying the correction method and uncertainty estimation to real experimental spectra.

## 3 EXPERIMENTAL

Our theoretical method was applied to measure and suitably correct a set of photometric spectra of samples consisting of various materials, including tetraphenyl-butadiene (TPB) films grown on glass substrate or dielectric reflecting foil (3M VM2000 [3]). TPB is the preferred material for the detection of ultraviolet (UV) radiation in modern particle detectors based on liquid Argon, especially for direct Dark Matter search [4,5]. Like other organic substances, it exhibits an intense broad emission band in the visible spectral interval, peaked at around 430 nm, under light excitation at shorter wavelengths [6]. Although some measurements in the Vacuum UV (VUV) range were performed in the past [7], in order to establish its conversion efficiency under excitation in the wavelength range 100-300 nm, a complete and systematic characterization in the optical spectral range is still missing.

The hemispheric transmittance and reflectance spectra of the samples listed in Table I were measured in the 250-600 nm wavelength range at room temperature (RT) by means of a

*Table I. List of names and descriptions of the analyzed samples.*

| Sample name | Description |
|---|---|
| VM2K_A | 3M VM2000 dielectric reflector foil [3]. |
| 2_F | Thin film consisting of TPB diluted in polystyrene over a glass substrate (1 mm thick). |
| CAMPIONE_1 | Surface layer of TPB over a VM2000 foil equal to sample VM2K_A. Naked weight = 0.151 g, TPB density ~600 μg/cm$^2$. |
| CAMPIONE_2 | Surface layer of TPB over a glass substrate (1 mm thick). Surface density = 1400 μg/cm$^2$. |
| CAMPIONE_5 | Surface layer of TPB over a glass substrate (1 mm thick). Surface density = 167 μg/cm$^2$. |

Perkin Elmer Lambda-19 spectrophotometer equipped with an integrating sphere of 150 mm in diameter. A Jobin Yvon Fluorolog-3 mod. FL-11 spectrofluorimeter was utilized to measure at RT the PL and PE spectra of the same samples in a front-face detecting geometry. The results are shown in Figs. 3 and 4, respectively. Note how, incidentally, the PL and PE spectra of two different materials like VM2000 and TPB almost fully overlap, thus allowing the use of a single filter for samples containing either VM2000 or TPB, or both. These measurements enabled us to select the most suitable optical filter among the available ones to apply our data correction procedure. The measured hemispherical transmittance and reflectance of the selected filter measured with the same spectrophotometer are shown in Fig. 5. The cut-off wavelength is $\lambda_0 \approx 390\,\text{nm}$ and at longer wavelengths the measured filter transmittance is instrumentally zero, with an uncertainty equal to the instrument sensitivity, that is, $\varepsilon = 0$ and $\delta\varepsilon = 5\times10^{-4}$ for the Perkin Elmer Lambda-19. Note that this filter is a band pass one rather than a low-pass one, its transmittance becoming zero again at wavelengths shorter than ~250 nm. This makes impossible to characterize the samples at those shorter wavelengths. Finally, Figure 6 shows the hemispherical transmittance and reflectance spectra of the glass substrate used for samples 2_F, CAMPIONE_2 and CAMPIONE_5.

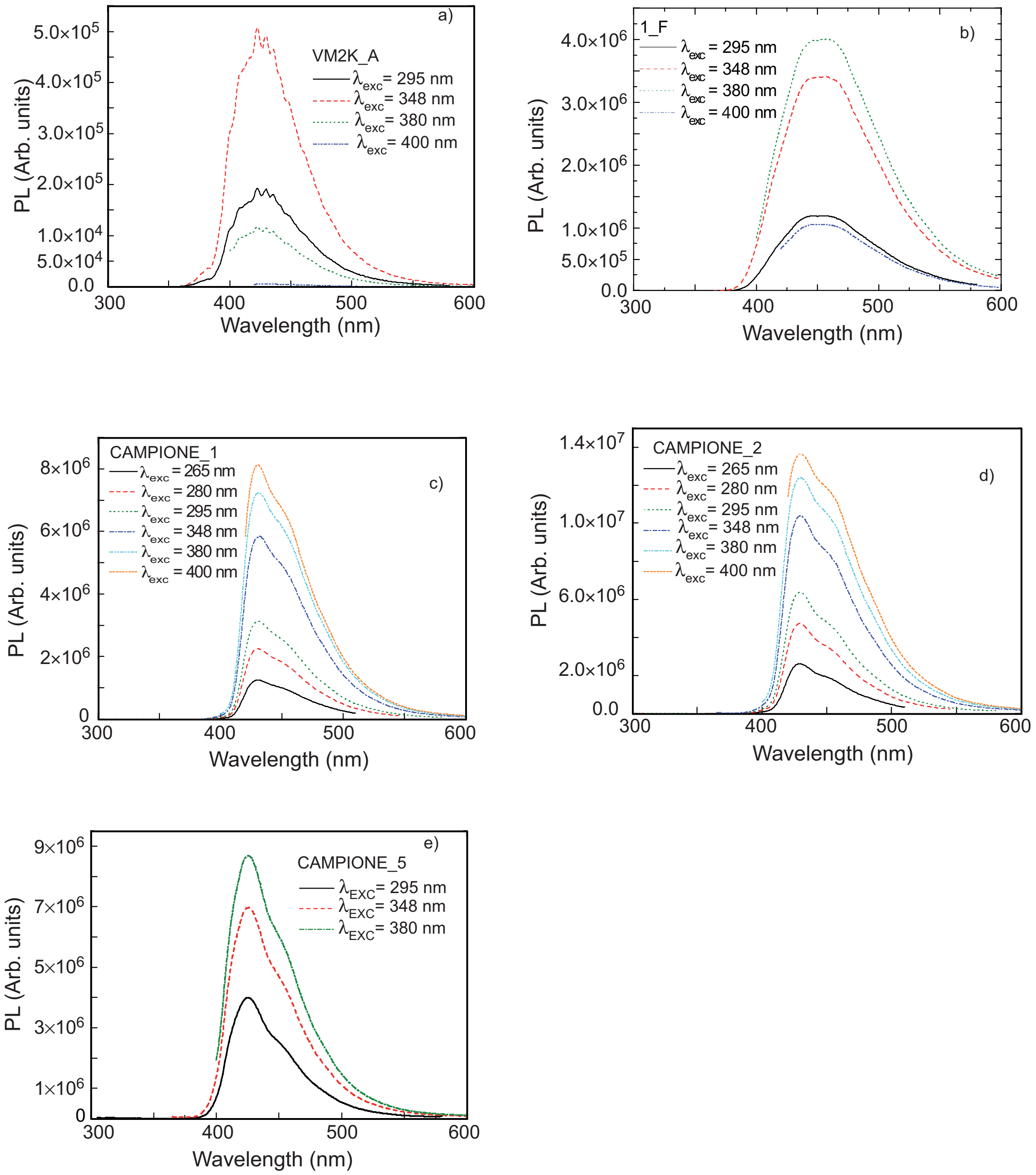

*Figure 3. Photoluminescence spectra of five samples. Sample 1_F was grown together with sample 2_F of Table I and should have similar properties. The other four samples are the same ones that are listed in Table I. The different lines correspond to different excitation wavelengths,* $\lambda_{exc}$ .

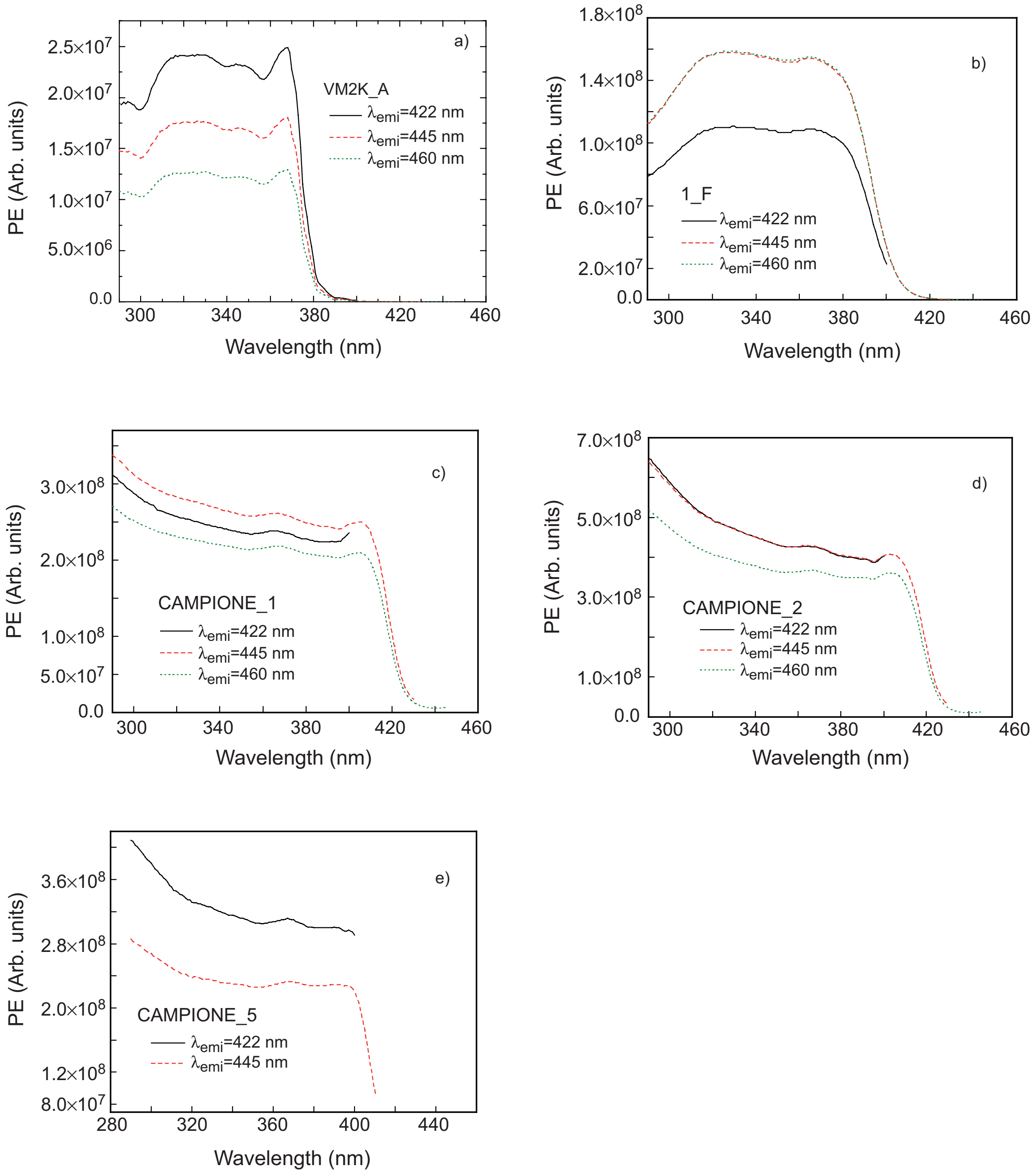

*Figure 4. Photoexcitation spectra of five samples. Sample 1_F was grown together with sample 2_F of Table I and should have similar properties. The other four samples are the same ones that are listed in Table I. The different lines correspond to different emission wavelengths,* $\lambda_{emi}$ .

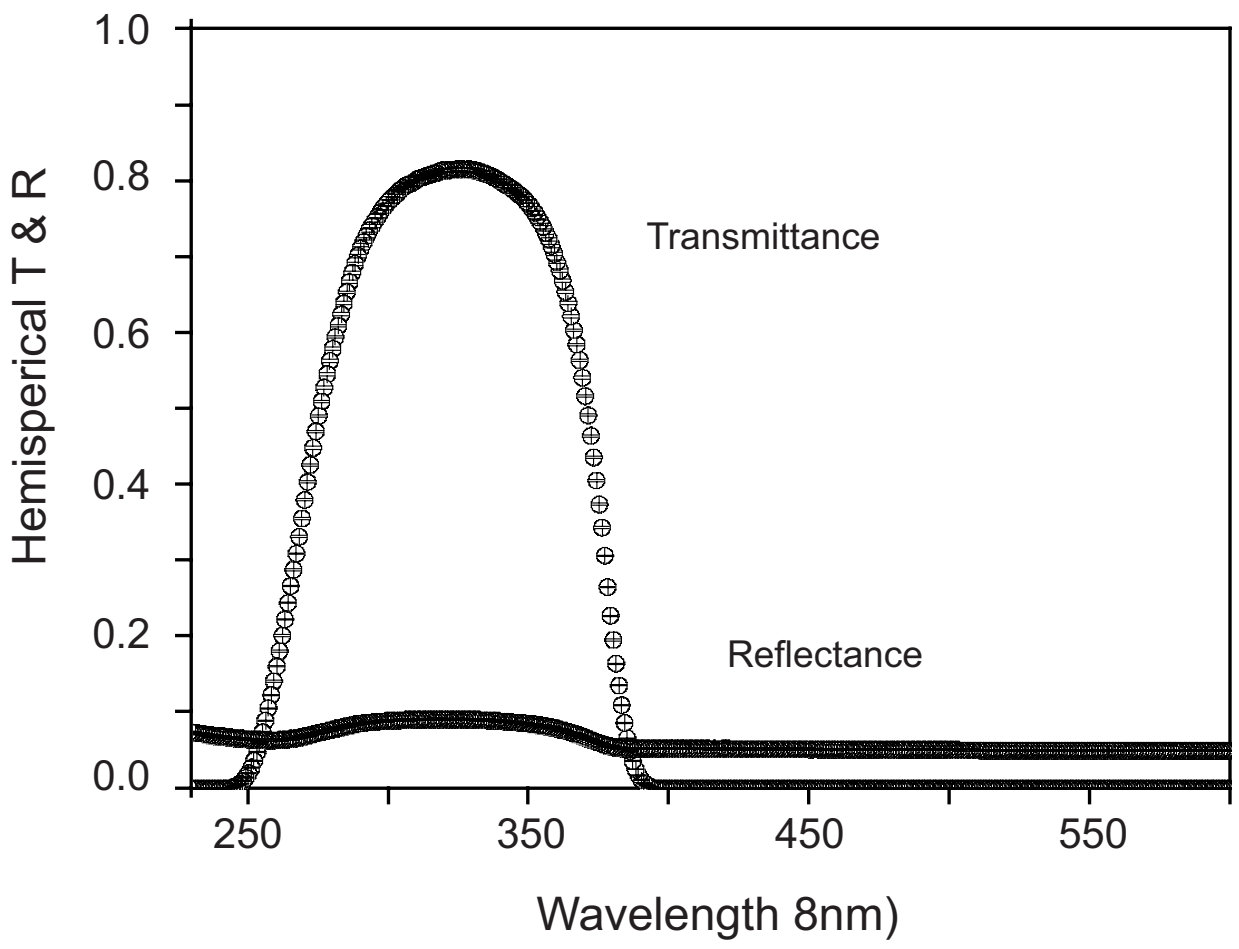

*Figure 5. Optical transmittance and reflectance of the low-pass filter utilized to correct the photometric spectra of the samples according to our theoretical approach.*

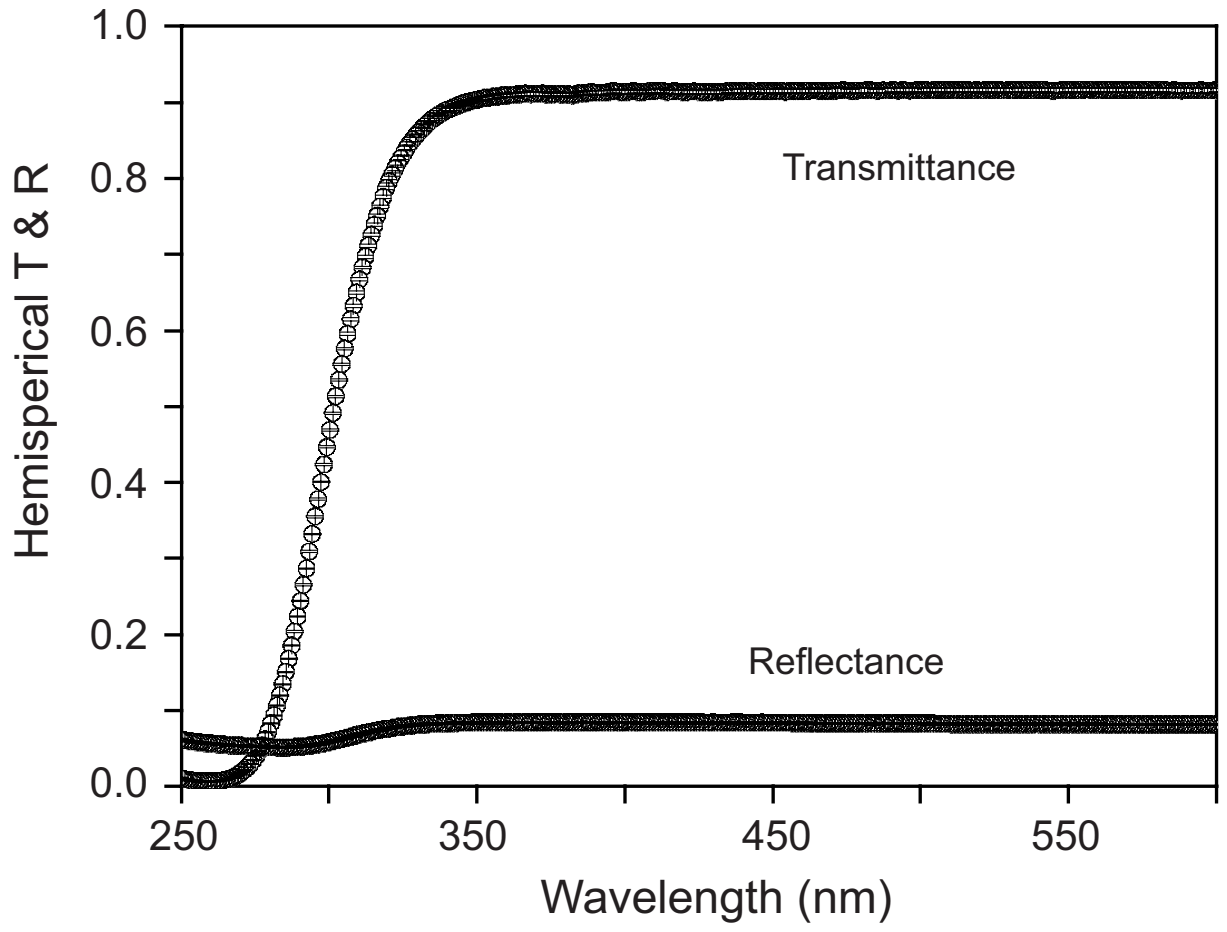

*Figure 6. Optical transmittance and reflectance of the glass substrate used for samples 2_F, CAMPIONE_2 and CAMPIONE_5.*

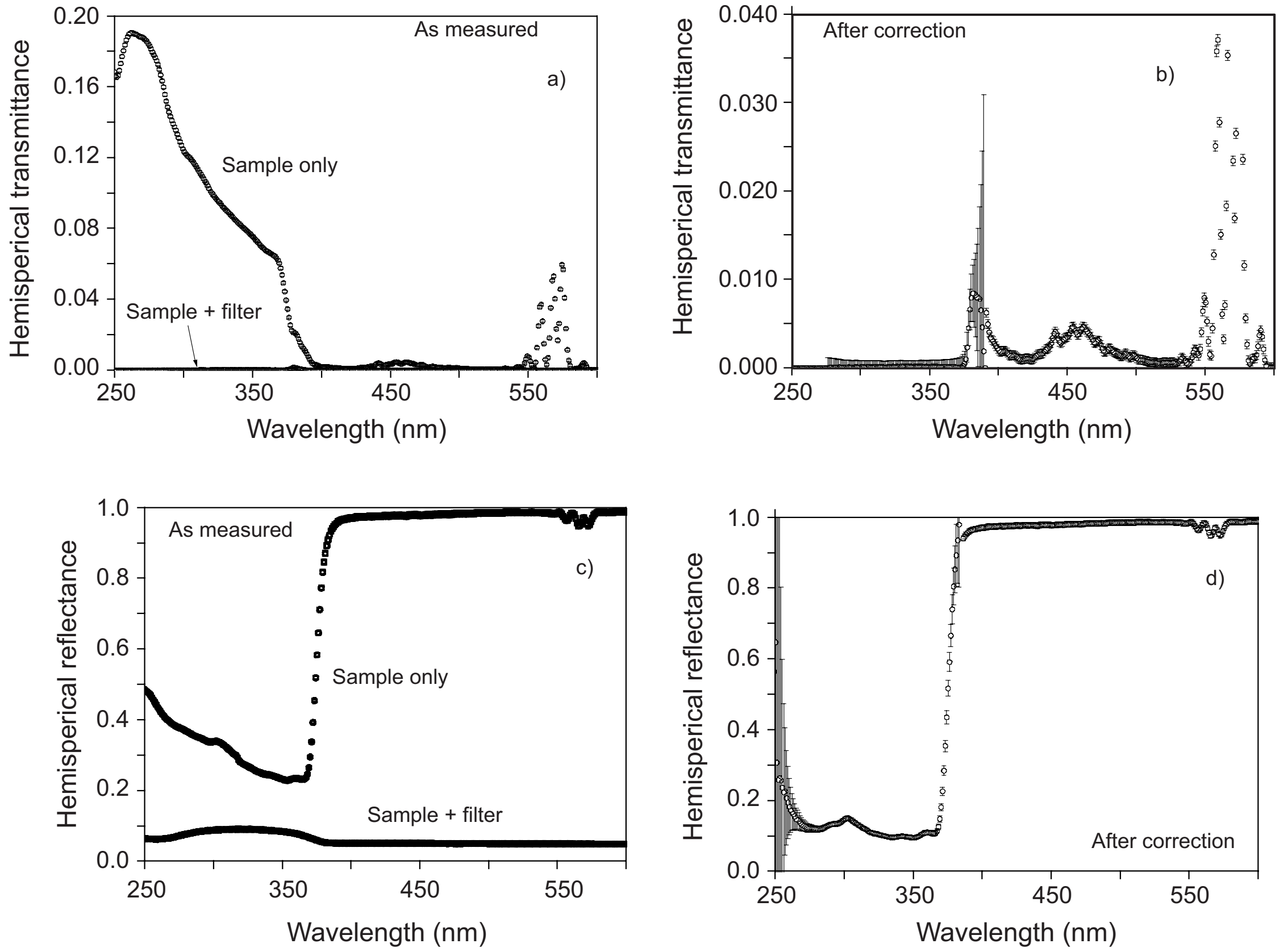

*Figure 7. As measured (including perturbation by PL) and corrected photometric spectra of sample VM2K_A: a) measured transmittance, b) corrected transmittance, c) measured reflectance, d) corrected reflectance. In a) and c) the transmittance and reflectance, respectively, of the assembly sample-filter are shown too.*

The as measured and corrected hemispherical reflectance and transmittance spectra of the samples in Table I are shown in Figs. 7-11. Note how for some those samples that have rather a dense content of TPB, see Figs. 9-11, the as measured curves assume values higher than 1 in the near ultraviolet. This would be of course absurd if only optically-passive properties of the samples were involved and is a signature of systematic errors that, as explained, we ascribe to unwanted PL contributions. As far as sample CAMPIONE_1 is concerned, see Fig. 9, only the reflectance spectra could be measured because transmittance was found to be instrumentally zero across all the examined spectral range. This fact could seem surprising, because the presence of two optically-active materials in the sample would induce to think that at least their PL should have been detected. However, there is an explanation. Sample CAMPIONE_1 consists of a layer of TPB over VM2000 and both of them are opaque in the visible range. The likely excited PL coming from the TPB layer was probably blocked by the VM2000 foil, while only a small amount of light within the PE band of VM2000 was able to pass through the TPB layer (and hence excite the VM2000 foil) and/or the VM2000 foil almost entirely absorbed the PL coming from itself.

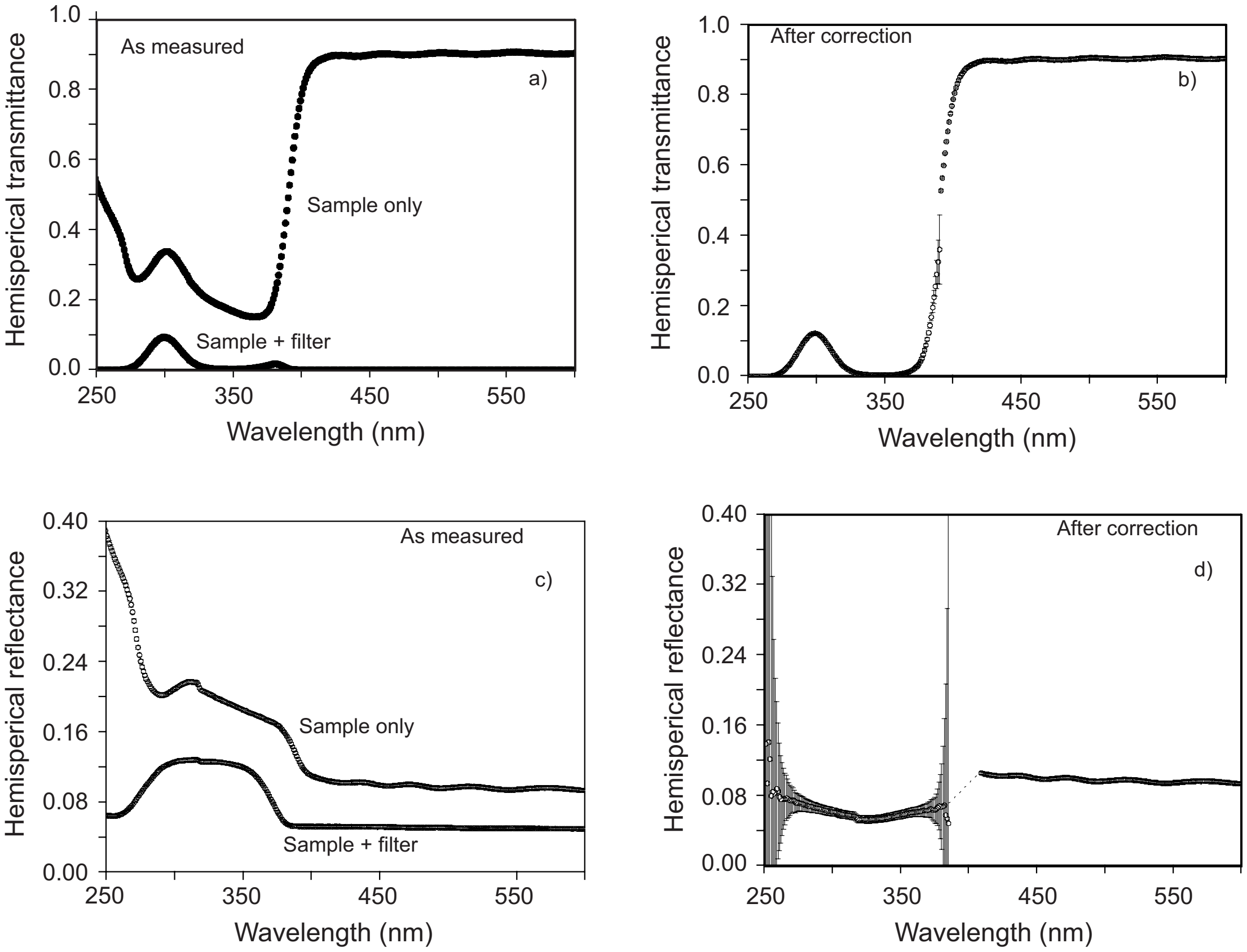

*Figure 8. As measured (including perturbation by PL) and corrected photometric spectra of sample 2_F: a) measured transmittance, b) corrected transmittance, c) measured reflectance, d) corrected reflectance. In a) and c) the transmittance and reflectance, respectively, of the assembly sample-filter are shown too.*

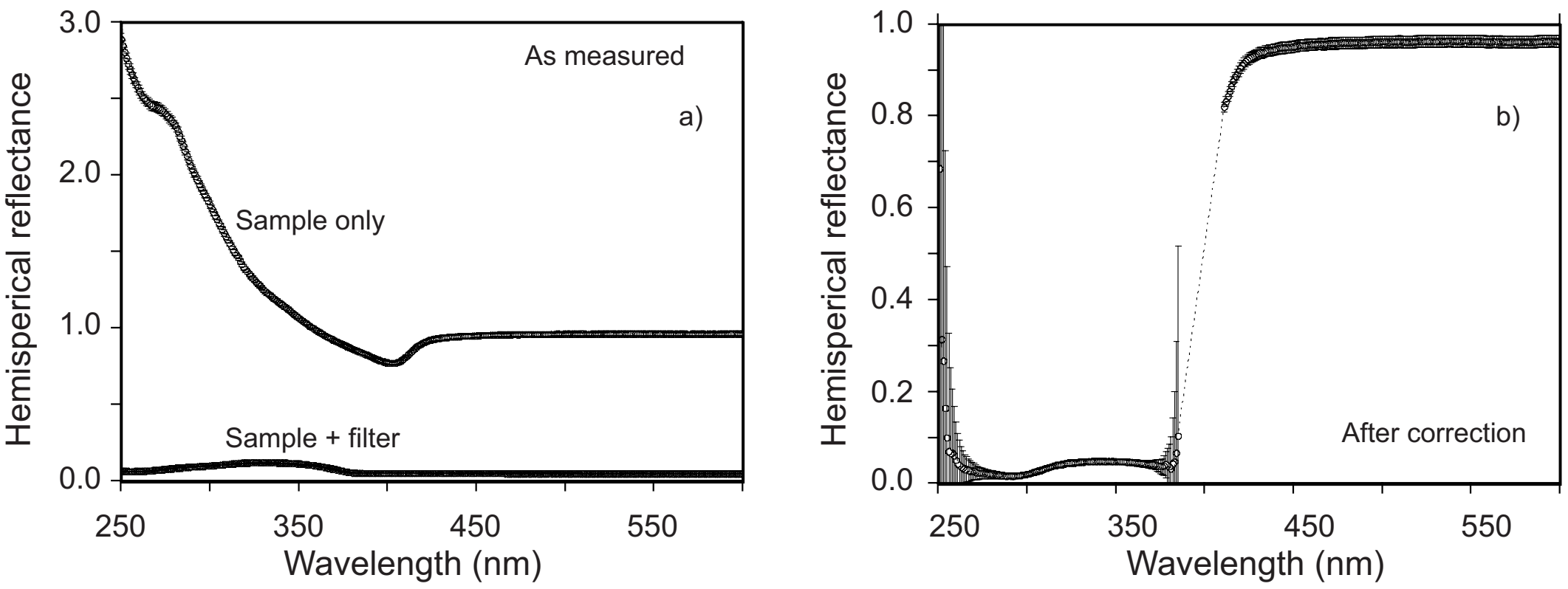

*Figure 9. As measured (including perturbation by PL) and corrected photometric spectra of sample CAMPIONE_1: a) measured reflectance, b) corrected reflectance. Transmittance measurement gave instrumental zero results over the entire analyzed spectrum due to the sample high opacity. In a) the reflectance of the assembly sample-filter are shown too.*

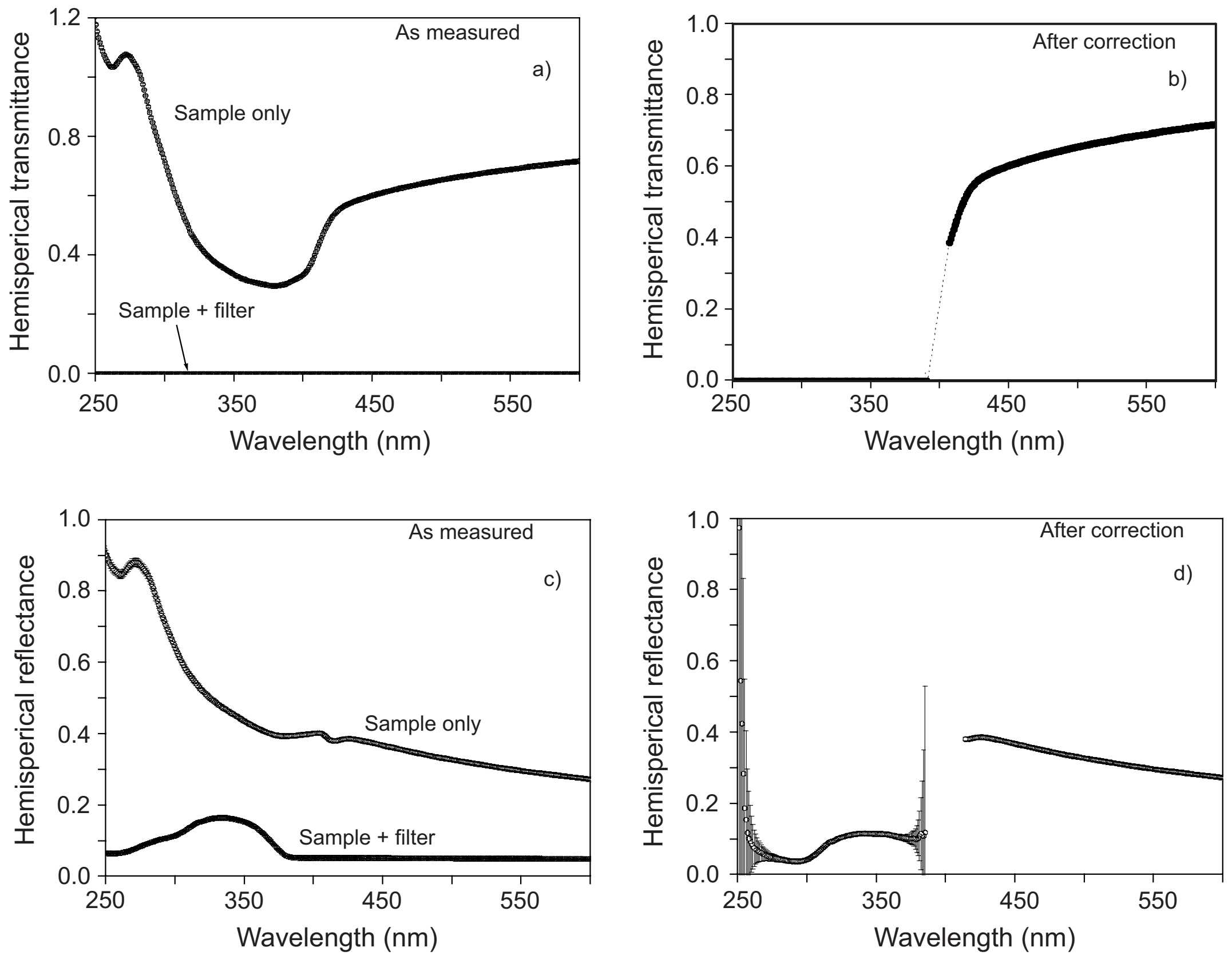

*Figure 10. As measured (including perturbation by PL) and corrected photometric spectra of sample CAMPIONE_2: a) measured transmittance, b) corrected transmittance, c) measured reflectance, d) corrected reflectance. In a) and c) the transmittance and reflectance, respectively, of the assembly sample-filter are shown too.*

## 4 DISCUSSION

The proposed data-correction method was tested with the samples listed in Table I. The results are shown in Figs. 7-11. Unfortunately, we were not able to cross-check the obtained results with spectra obtained by different means. To do that, a possible approach could be measuring the *direct* transmittances and *specular* reflectances of the samples in a spectrophotometer, knowing that for these kinds of measurements the perturbation due to PL should be much smaller than in presence of an integrating sphere thanks to the smaller detection angle, and then compare the obtained spectra with the corrected ones in Figs. 7-11. However, because of the strong scattering nature of the analyzed samples, the comparison would be affected by the lack of detection of scattered light in the direct and specular measurements, hence it makes no sense.

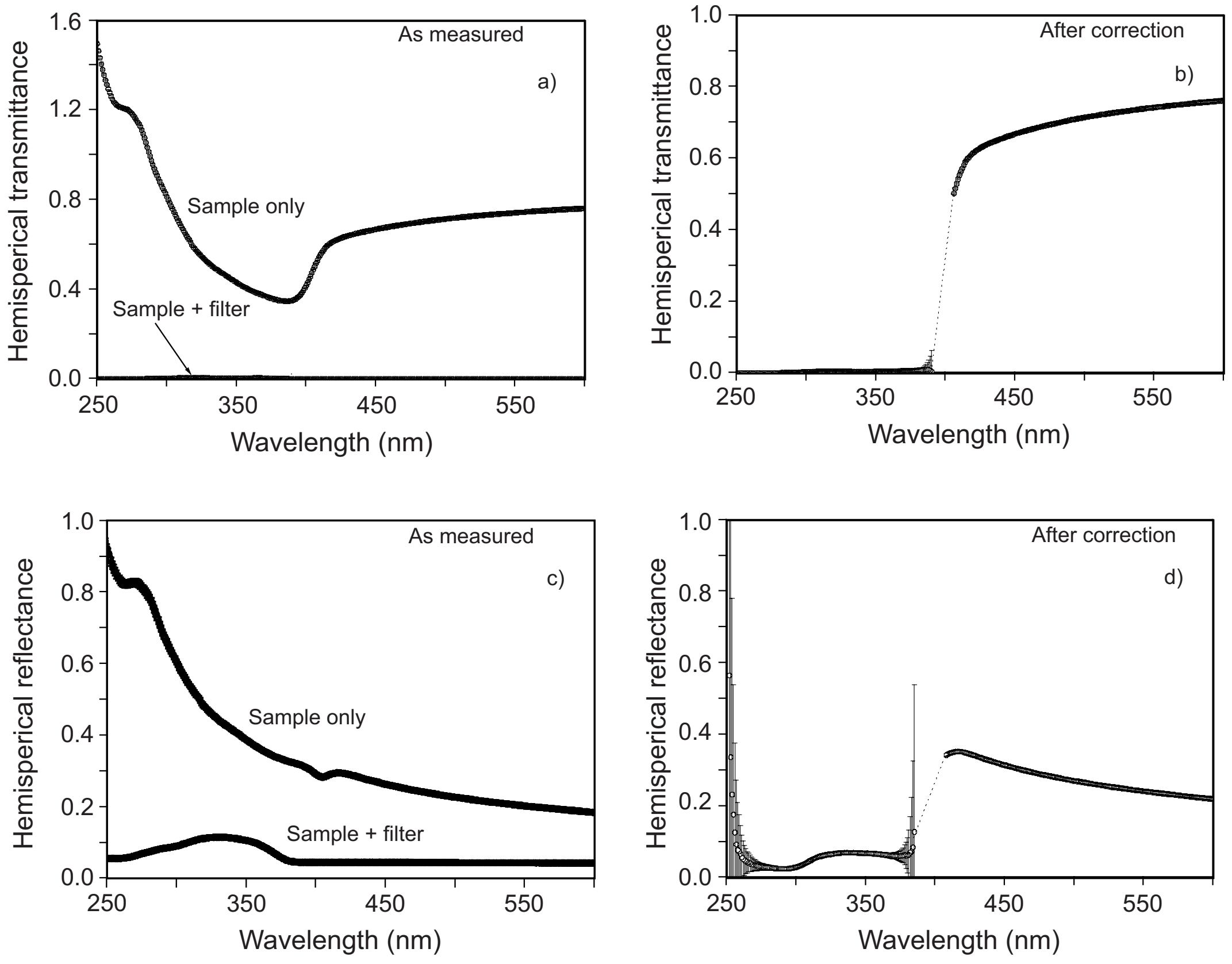

*Figure 11. As measured (including perturbation by PL) and corrected photometric spectra of sample CAMPIONE_5: a) measured transmittance, b) corrected transmittance, c) measured reflectance, d) corrected reflectance. In a) and c) the transmittance and reflectance, respectively, of the assembly sample-filter are shown too.*

All the examined samples approximately share the same wavelength ranges as far as PL and PE bands are concerned. This fact simplified the choice of a unique suitable low-pass filter for the additional spectral measurements of the whole set that were later used to process and correct the original measurements according to our theory. If this was not the case, distinct optical filters should have been chosen for the task.

Regarding the samples consisting of layers over glass substrates, Fig. 6 shows that the glass band gap is located at about 320 nm and that at wavelengths shorter than 270 nm the substrate transmittance is practically zero. The overall transmittance of samples 2_F, CAMPIONE_2 and CAMPIONE_5 at those wavelengths is therefore very close to the instrumental zero because of the substrate contribution. Incidentally, the small transmittance of the selected filter at wavelengths around 250 nm and shorter can make the data-correction uncertainties quite large at those wavelengths for the appearance of the term $T_F - \varepsilon$ in the denominators of Eqs. (11) and (12). For our samples, this has been found to be especially true for reflectance, as witnessed by the large error bars in Figs. 7d)-11d).

As far as the aforementioned "gray zone" of the spectrum is concerned, approximately enclosed between the filter cut-off wavelength, $\lambda_0$, and the lower limit of the PL band and wherein the data-correction uncertainties could diverge, we adopted, when needed, a linear interpolation between the useful spectral distribution tails, which can be seen as dotted lines in Figs. 8-11. We have not found a better way to tackle data correction in this spectral zone.

## 5 CONCLUSIONS

Measuring the photometric spectra of optically-active samples with a commercial grade spectrophotometer can be rather a difficult task because luminescence induced by PE can be generated by the sample and reach the instrument detector, thus perturbing the measurement. This is particularly true if an integrating sphere is utilized, because generally PL is spread over all the solid angle and thus almost entirely captured by the detecting system.

In this Report we have developed a method to correct PL-perturbed photometric spectra by means of additional measurements performed with a suitable optical filter, which has to be selected on the basis of PL and PE measurements. Although the filter can be very efficient in eliminating the PL contribution arriving to the detector, it also perturbs the measurement within the PE band because of its optical transmittance and reflectance, which are different from an ideal optical response. Nonetheless, a proper elaboration of the spectra taken without and with the filter can lead to a reliable estimation of the *true* optical characteristics of the examined sample, even though within certain intervals the high uncertainty of the estimation makes a linear interpolation approach more feasible.

The elaboration approach introduced in this Report was applied to five samples containing photoluminescent materials such as TPB and VM 2000. The incidental fairly-good superposition of the PL and PE bands of these two materials made the filter selection process easier, especially as far as the sample containing both of these materials (CAMPIONE_1) is concerned.

## ACKNOWLEDGEMENTS

Research carried out within TECVIM project *Tecnologie per Sistemi di Visualizzazione di Immagini*, funded by the Italian Ministry of University and Scientific Research MIUR as support to applied research, and within the Research Contract *Studio di Film Organici Emettitori di Luce per Applicazioni nei Rivelatori di Radiazione mediante Misure Spettro-Fotometriche*, funded by INFN-LNGS, Istituto Nazionale di Fisica Nucleare – Laboratori Nazionali del Gran Sasso. All the samples in use with this test have been produced at INFN Gran Sasso by means of a dedicated Knudsen-type vacuum evaporation set-up developed and realized by INFN-Pavia.